\documentclass[11pt]{article}

\usepackage{moriond,epsfig}
\usepackage[latin1]{inputenc}
\usepackage[OT1]{fontenc}

\def\Journal#1#2#3#4{{#1} {\bf #2}, #3 (#4)}
\def\Article#1#2#3#4{{#1} {\bf #2}, (#3) #4}
\def\Proc#1#2{{#1}, (#2)}

\def\NIM{\em Nucl. Instrum. Methods Phys. Res.}
\def\AP{\em Astropart. Phys.}
\def\HPH{\em F. Helv. Phys. Acta}
\def\LTD12{J.Low Temp. Phys.: Proceedings of the 12$^{{\it th}}$ International Workshop Low Temperatures Detectors}
\def\NIMA{{\em Nucl. Instrum. Methods} A}
\def\NPB{{\em Nucl. Phys.} B}

\def\PRD{{\em Phys. Rev.} D}

\def\be{\begin{equation}}
\def\ee{\end{equation}}
\def\bea{\begin{eqnarray}}
\def\eea{\end{eqnarray}}

\begin{document}
\vspace*{4cm}
\title{THE EDELWEISS-II EXPERIMENT}

\author{ S.~SCORZA, for the EDELWEISS collaboration}

\address{Institut de Physique Nucl{\'e}aire de Lyon, Universit{\'e} Lyon 1, Universit{\'e} Claude Bernard Lyon 1, CNRS/IN2P3, 4,Rue Enrico Fermi, 69622 Villeurbanne Cedex, France}

\maketitle\abstracts{EDELWEISS is a direct dark matter search situated in the low radioactivity environment
of the Modane Underground Laboratory.
The experiment uses Ge detectors at very low temperature in order to identify
eventual rare nuclear recoils induced by elastic scattering of WIMPs from
our Galactic halo.
The commissioning of the second phase of the experiment, involving more than
7 kg of Ge, has been completed in 2007.
Two new type of detectors with active rejection of events due to surface contamination
have demonstrated the performances required to achieve the physics goal of
the present phase.}

\section{Introduction}
First indications for the existence of Dark Matter were already found in the 1930s~\cite{zw}. By now there is strong evidence~\cite{gbdhjs} to believe that a large fraction (more than 80\%) of all matter in the Universe is Dark (interacts very weakly with electromagnetic radiation, if at all) and that this Dark Matter is predominantly non-baryonic.
Weakly Interacting Massive Particles (WIMPs) are one of the leading candidates for Dark Matter. WIMPs are stable particles which arise in several extensions of the Standard Model of electroweak interactions~\cite{hhnos}.
Typically they are presumed to have masses between few tens and few hundreds of GeV/c$^{2}$ and a scattering cross  section with a nucleon below 10$^{-6}$ pb.

The EDELWEISS experiment (Exp{\'e}rience pour D{\'e}tecter les WIMPs en Site Souterrain) is dedicated to the direct detection of WIMPs. The direct detection principle consists in the measurement  of the energy released by nuclear recoils produced in an ordinary matter target by the elastic collision of a WIMP from the Galactic halo. The main challenge is the expected extremely low event rate ($\leq$ 1evt/kg/year) due to the very small interaction cross section of WIMP with the ordinary matter. An other constraint is the relatively small deposited energy ($\leq$ 100keV).

\section{Experimental setup and detectors}
The EDELWEISS experiment is situated in the Modane Underground Laboratory (LSM) in the Fr{\'e}jus highway tunnel, where an overburden of about 1700 m of rock, equivalent to 4800 m of water, reduces the cosmic muon flux down to 4.5~$\mu$/m$^{2}$/day, that is about 10$^{6}$ times less than at the surface. The installation of the second phase of the experiment, EDELWEISS-II, was completed beginning of 2006. In order to measure low energy recoils, EDELWEISS employs cryogenic detectors (high purity Ge crystal) working at temperature of about 20 mK, with simultaneous measurements of phonon and ionization signal. The ionization signal, corresponding to the collection on electrodes of electron-hole pairs created by the energy loss process, depends on the particle type whereas the heat signal reflects the total energy deposit. This simultaneous measurements of two signals allows an event by event discrimination between the electronic recoils, tracers of background (induced by photons and electrons) and the nuclear one originated by neutrons and WIMPs.

By reducing the radioactive background and increasing the detection mass, EDELWEISS should reach cross-section sensitivities of interest for SUSY models. Specific improvements are aimed at reducing the possible background sources~\cite{sf} that have limited the sensitivity of EDELWEISS-I~\cite{vs}.

\begin{figure}
\begin{center}
\psfig{figure=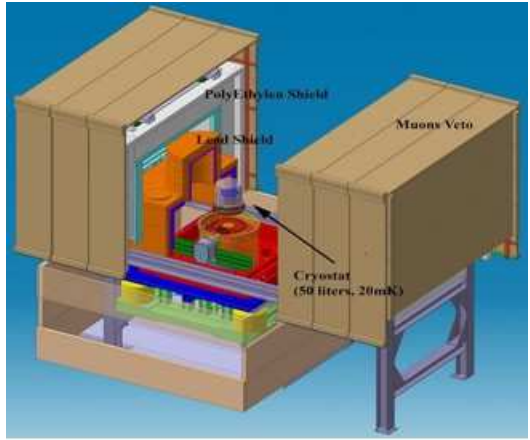,height=2.3in}
\caption{General scheme of the EDELWEISS-II experiment. From Outside to inside: The outer shell is the muon veto system, followed by the polyethylene shield and the inner lead shielding. The upper part can be open to have access to the cryostat  which houses the bolometers.
\label{fig:scheme}}
\end{center}
\end{figure}

To reduce environmental background, all materials used in vicinity of the detectors have been tested for their radiopurity with a dedicated HPGe detectors.
A class 10$^{4}$ clean room surrounds the whole setup shown in Fig.~\ref{fig:scheme}. A class 100 laminar flow with deradonised air ($\leq$ 0.1Bq/m$^{3}$) is used when mounting the detectors in the cryostat.
The gamma background is screened by a 20~cm thick lead shielding around the cryostat. Nuclear recoils induced by neutrons constitue an ultimate physical background. In the experimental volume of the cryostat the neutron background is attenuated by three orders of magnitude thanks to a 50~cm thick polyethylene shielding. In addition, an active muon veto with a coverage of more that 98\% tags the muons interacting in the lead shield and producing neutrons.
The residual neutron background comes from high energy neutrons produced by muons that are not tagged by the veto system and by neutrons from $^{238}$U fission in the lead shielding. 
By Monte Carlo simulations the nuclear recoil rate above 10~keV in the detectors is estimed to be $<$10$^{-3}$~evt/kg/d, that corresponds to a WIMP-nucleon cross-section sensitivity of 10$^{-8}$~pb for a WIMP mass of $\sim$100~GeV/c$^{2}$, improving the sensitivity of a factor 100 compared to EDELWEISS-I, and competitive with CDMS-II~\cite{cdms2}.

The dilution cryostat is of a reversed design, with the experimental chamber on the top of the structure. Its large volume (50~l) can host up to 120 detectors arranged in a compact way, that will allow increasing the number of neutron coincidences.

One of the main sources of background for EDELWEISS are the surface events. When the interaction takes place near the electrodes, the charge collection can be incomplete, resulting in ionization signals smaller than the expected ones that can mimic nuclear recoils.
Besides the 320~g Ge/NTD detectors type of EDELWEISS-I, equipped with holders, two other detection techniques for active surface events rejection have been tested: the 400~g Ge/NbSi type, replacing the NTD thermistors by two NbSi Anderson insulator thermometric layers sensitive to athermal phonons~\cite{aj}$^{,}$~\cite{sm} and the 400~g Ge/NTD crystal with a special interdigitized electrodes scheme recently developped~\cite{abr}$^{,}$~\cite{xd}.
The surface event rejection capabilities of Ge/NbSi and Ge/NTD with interdigitized electrodes have been measured to be better than 95\%~\cite{sm}$^{,}$~\cite{xd} for events occurring in the first millimeters under the detector surface.

\section{Commissioning run} 
The year 2006 has been dedicated to the tuning of the electronics and improvements on the acquisition and cryogenics, especially for acoustic and mechanical decoupling of the thermal machines, performing four cryogenic runs with 8 bolometers. 
In 2007, commissioning runs were performed with 21 Ge/NTD and 4Ge/NbSi. The plan in the next years is to add more detectors with surface event rejection capabilities.

\begin{figure}
\begin{center}
\psfig{figure=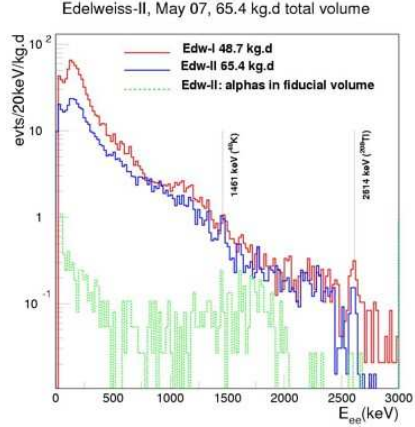,height=2.3in}
\caption{Ionization energy spectrum for low background run for the Ge/NTD detectors (65kg.d for the total volume). The comparison with EDELWEISS-I gamma background shows an improvementof a factor 2 at lox energy. The alpha background are identified by selecting events with a ionization/recoil energy ratio less than 0.5 and recoil energy between 3~MeV and 8~MeV.  
\label{fig:gamma}}
\end{center}
\end{figure}

In addition to gamma and neutron source calibration, low background runs have been performed to quantify the alpha, gamma and beta backgrounds. The aim of these runs is to measure the rate of surface events with incomplete charge collection in the detectors to extrapolate to the future sensitivity. The electronic recoil background rate for the fiducial volume~\cite{om} is shown in Fig.~\ref{fig:gamma}. For energy below 100~keV, it is approximately 0.6~evt/kg/d which is a factor 2.5 better than EDELWEISS-I. The alpha background is between 1.6 and 4.4~$\alpha$/kg/d (75-200~$\alpha$/m/d), depending on the detector and its near-by environment. The mean alpha rate for EDELWEISS-I was 4.2~$\alpha$/kg/d.

\begin{figure}
\begin{center}
\psfig{figure=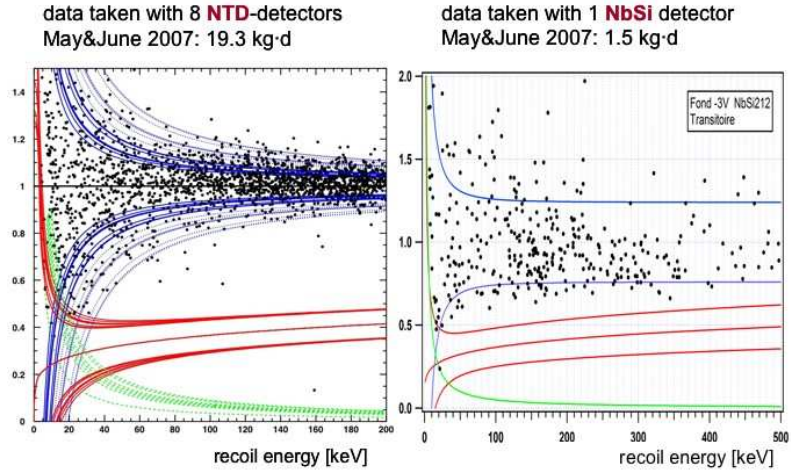,height=2.5in}
\caption{Distribution of the quenching factor (ionization/recoil energy ratio) as a function of the recoil energy for data collected in background runs - Left: for 8 Ge/NTD detectors, Right: for one Ge/NbSi detector.
\label{fig:qplot}}
\end{center}
\end{figure}

Results of low background commissioning run are shown in Fig.~\ref{fig:qplot}.
The left hand side shows the result corresponding to a fiducial exposure of 19.3~kg.d of the 8 Ge/NTD detectors with lowest threshold. No events are observed in the nuclear recoil band showing a possible improvement compared to EDELWEISS-I. More statistics, acquired in stable and well-defined conditions, are obviously needed to draw firm conclusion.  
Results obtained for a 200~g Nbsi for a fiducial exposure of 1.5~kg.d after the cuts removing the surface events are shown in the right hand side. The resolutions need tuning: the rather large dispersion centered at Q=1.0 is due to discrimination parameter obtained with athermal signal.

\section{Conclusions and prospects}

The EDELWEISS-II setup has been validated with calibration and background runs.
Energy resolution and discrimination capabilities close to those of EDELWEISS-I have been measured for Ge/NTD detectors.
The validation of Ge/NbSi detectors with new aluminum electrodes is in progress and Ge/NTD detectors with interdigitized electrode scheme have shown promising results in surface laboratory.

Low background physics runs will be taken with 28 detectors setup with the aim to reach sensitivity to WIMP-nucleon cross-section of 10$^{-7}$~pb for a WIMP mass of 100~GeV in 2008.
Additional detectors will be added in the two coming years to enhance progressively the sensitivity to few 10$^{-9}$~pb thanks to active surface rejection.

\end{document}